\pdfoutput=1
\documentclass[amsmath,amssymb,aps,twocolumn,superscriptaddress]{revtex4-2}
\usepackage{amsmath}
\usepackage{amssymb}
\usepackage{bm}
\usepackage{graphicx}
\usepackage{epsf}
\usepackage{xcolor}
\usepackage{color}
\usepackage{lipsum}
\usepackage{xr}
\usepackage[export]{adjustbox}
\usepackage[version=4]{mhchem}
\usepackage[bookmarks=false]{hyperref}
\usepackage{siunitx}
\usepackage[english]{babel}
\usepackage[T1]{fontenc}
\usepackage{float}
\usepackage{dcolumn}
\usepackage{bibunits}
\makeatletter
\def\maketitle{
\@author@finish
\title@column\titleblock@produce
\suppressfloats[t]}
\makeatother

\begin{document}
\title{Counterdirectional Exciton and Trion Motion in Applied Electric Field}
	
\author{Daniel Vaz}
\thanks{These authors contributed equally to this work}
\affiliation{Department of Physics, University of Pittsburgh, 3941 O'Hara Street, Pittsburgh, Pennsylvania 15218, USA}

\author{Yuanjun Guan}
\thanks{These authors contributed equally to this work}
\affiliation{State Key Laboratory of Precision Spectroscopy, East China Normal University, Shanghai 200241, China}
    
\author{Qiaochu Wan}
\affiliation{Department of Physics, University of Pittsburgh, 3941 O'Hara Street, Pittsburgh, Pennsylvania 15218, USA}
	
\author{Bobby Bobby}
\affiliation{Department of Physics, University of Pittsburgh, 3941 O'Hara Street, Pittsburgh, Pennsylvania 15218, USA}
	
\author{Anshul Ramavath}
\affiliation{Department of Physics, University of Pittsburgh, 3941 O'Hara Street, Pittsburgh, Pennsylvania 15218, USA}
	
\author{Brandon Vargo}
\affiliation{Department of Physics, University of Pittsburgh, 3941 O'Hara Street, Pittsburgh, Pennsylvania 15218, USA}
	
\author{Juntong Ye}
\affiliation{Department of Physics, University of Pittsburgh, 3941 O'Hara Street, Pittsburgh, Pennsylvania 15218, USA}

\author{Yuzhi He}
\affiliation{Department of Electrical and Computer Engineering, University of Pittsburgh, 3941 O'Hara Street, Pittsburgh, 15212, PA, USA}
    
\author{Jonathan Beaumariage}
\affiliation{Department of Physics, University of Pittsburgh, 3941 O'Hara Street, Pittsburgh, Pennsylvania 15218, USA}

\author{Om Patel} 
\affiliation{Sharpsville Area Senior High School, 301 Blue Devil Way, Sharpsville, PA 16150, USA}
	
\author{Kenji Watanabe}
\affiliation{Research Center for Electronic and Optical Materials, National Institute for Materials Science, 1-1 Namiki, Tsukuba, 3050044, Ibaraki, Japan}
	
\author{Takashi Taniguchi}
\affiliation{Research Center for Materials Nanoarchitectonics (MANA), National Institute for Materials Science, 1-1 Namiki, Tsukuba, 3050044, Ibaraki, Japan}

\author{Xiong Feng}
\affiliation{Department of Electrical and Computer Engineering, University of Pittsburgh, 3941 O'Hara Street, Pittsburgh, 15212, PA, USA}

\author{James Hone}
\affiliation{Department of Mechanical Engineering, Columbia University, 500 W. 120th Street, New York, 10027, NY, USA}
	
\author{Nathan Youngblood}
\affiliation{Department of Electrical and Computer Engineering, University of Pittsburgh, 3941 O'Hara Street, Pittsburgh, 15212, PA, USA}
	
\author{Zheng Sun}
\thanks{Address correspondence to: zsun@lps.ecnu.edu.cn}
\affiliation{State Key Laboratory of Precision Spectroscopy, East China Normal University, Shanghai 200241, China}
\affiliation{Shanghai Key Laboratory of Magnetic Resonance, School of Physics and Electronic Science, East China Normal University, Shanghai 200241, China.}
\affiliation{Collaborative Innovation Center of Extreme Optics, Shanxi University, Taiyuan, Shanxi 030006, China.}
	
\author{David W. Snoke}
\thanks{Address correspondence to: snoke@pitt.edu}
\affiliation{Department of Physics, University of Pittsburgh, 3941 O'Hara Street, Pittsburgh, Pennsylvania 15218, USA}

\date{\today}
\begin{bibunit}[apsrev4-2]

\begin{abstract}
Charged excitonic complexes are central to the optoelectronic and many-body properties of semiconductors, yet their real-space transport dynamics remain largely unexplored. Here, we report the direct optical observation of trion motion under an applied electric field. The trions exhibit electrically driven drift with velocities approaching {$10^5~\mathrm{m/s}$}. Unexpectedly, the trion flow induces a pronounced back-action on coexisting neutral excitons, driving them in the opposite direction and giving rise to counterpropagating exciton-trion transport. Our results reveal an interaction-driven nonequilibrium transport regime of mixed excitonic fluids and establish a direct route for imaging the dynamics of more complex charged quasiparticles, including doubly charged excitons.
\end{abstract}

\maketitle

\section{Introduction}

Trions, or charged excitons, are three-particle bound states formed when an exciton (electron-hole pair) binds to an additional charge carrier. As trions carry both charge and optical coherence, they provide a unique platform for exploring the interplay between light-matter interactions~\cite{Dhara2018,Paik2023,Guan2025} and electrically driven quasiparticle transport~\cite{Cheng2021,lee2023electric}. Since their first observation in modulation-doped GaAs quantum wells more than three decades ago~\cite{kheng1993observation}, trions have become a basic element of many-body physics in semiconductors, providing a platform for investigating Coulomb correlations~\cite{Mak2013} and exciton-carrier interactions~\cite{Ross2013}, as well as the crossover between trion and Fermi-polaron descriptions in doped systems~\cite{Efimkin2017,Sidler2017,liu2021exciton}.

Subsequent experiments demonstrated that trions can be transported by externally applied electric fields, revealing their charged-particle nature and opening opportunities for electrically controlled excitonic devices~\cite{sanvitto2001observation,pulizzi2003optical}. However, the relatively small binding energies of excitons and trions in conventional III-V semiconductors limit such studies to cryogenic temperatures.

By contrast, the emergence of atomically thin transition-metal dichalcogenides (TMDCs) has transformed this landscape. Owing to reduced dielectric screening and strong Coulomb interactions, monolayer TMDCs host tightly bound trions with binding energies exceeding 20--30$~\mathrm{meV}$~\cite{mak2013tightly,Courtade2017,wang2018colloquium}, allowing stability even at room temperature. As a result, trionic resonances have emerged as a defining feature of the optical response of doped TMDC monolayers and heterostructures.  Beyond their fundamental importance, the charged nature of trions enables direct coupling to electric fields while preserving strong optical interactions, making them attractive for electrically controllable excitonic and valleytronic functionalities~\cite{Ross2013,Mak2014}. Their charged nature further enables interactions with electric fields, carrier reservoirs and other excitonic species, giving rise to transport phenomena that have no direct analog in conventional photonic systems.

Prior studies of trions in 2D materials have reported trion drift in response to applied field~\cite{lee2022drift, lee2023electric, Cheng2021,Kato2016}, but the mobility of the trions limited the resolution of the measurements, and it could also be hard to distinguish motion due to electric field from motion induced by strain gradients. More fundamentally, it remains largely unexplored how the motion of charged excitonic complexes influences coexisting neutral excitons within the same many-body system.

Here we present results on free trion drift in encapsulated monolayer MoSe$_2$ using two complementary device architectures. In one type, an hBN-encapsulated MoSe$_2$ monolayer with graphene contacts and a bottom-gate geometry enables independent electrostatic doping and the application of in-plane electric fields, enabling quantitative measurements of field-driven trion transport. In a second set, time-resolved spatial imaging was performed in a capacitor-like MoSe$_2$/hBN heterostructure, providing direct visualization of excitonic motion. Despite their distinct architectures and measurement approaches, both platforms reveal consistent transport behaviour.

Under the applied field, negative trions drift opposite to the in-plane field direction, as expected from their charge. Remarkably, the trion motion simultaneously induces a pronounced back-action on coexisting neutral excitons, driving them in the opposite direction and giving rise to a counterpropagating exciton-trion flow. This observation suggests that transport-induced trion density gradients can exert a substantial influence on the dynamics of nearby excitons, revealing an unexpected form of coupling between neutral and charged excitonic populations. To elucidate the underlying mechanism, we develop a coupled drift-diffusion model for exciton and trion transport. The theoretical results reproduce the experimentally observed linear field dependence of the transport for both quasiparticle species, showing excellent agreement with the measurements.

\section{Sample fabrication and measurement setup}

\begin{figure}[!htb]
\begin{center} 
\includegraphics[width=\linewidth]{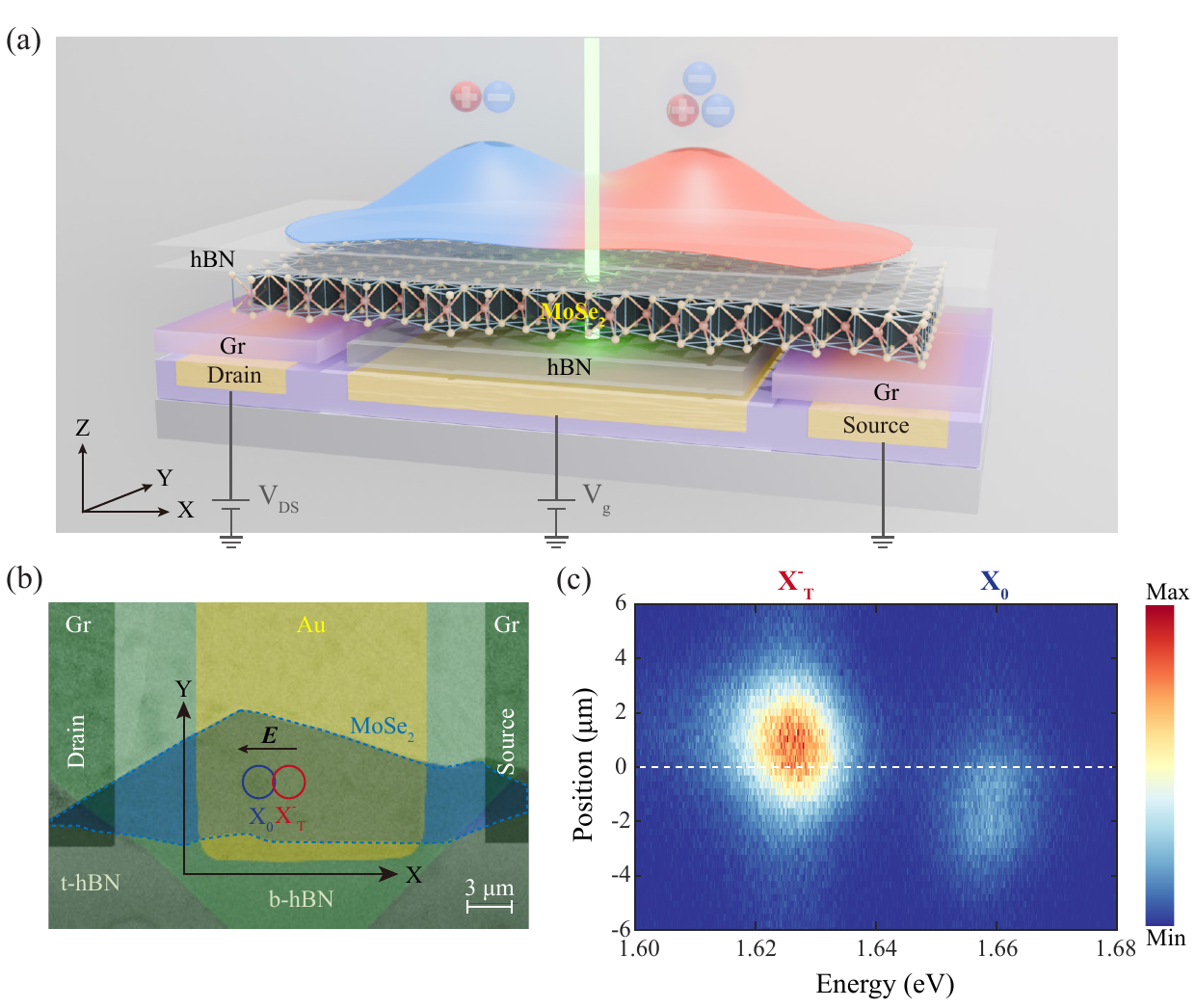}
\end{center}
\caption{Architecture of Device I and characterization of the sample used for electrically driven trion transport measurements.(a) Schematic of the device architecture. $V_\text{DS}$ is the voltage applied between the source and drain graphene contacts to generate the in-plane electric field, while $V_\text{G}$ is the voltage applied between the back gate and the source graphene gate to tune the doping density of the MoS$_2$ monolayer. (b) Optical micrograph of the device. All graphene contacts are connected to Au wires and subsequently to Au bonding pads for electrical measurements. (c) Representative spatially resolved PL spectrum acquired from the region indicated in (b). The MoSe$_2$ is intrinsically $n$-doped, making the negative trion the dominant excitonic species. Under an applied in-plane electric field, negative trions drift opposite to the field direction.}
\label{fig1}
\end{figure}

Device I consisted of an hBN-encapsulated monolayer MoSe$_2$ on a SiO$_2$/Si substrate, integrated with pre-patterned graphene contacts and an Au/Ti back gate, as illustrated in Figure~\ref{fig1}(a). The Au back gate was deposited after locally etching the SiO$_2$/Si substrate to achieve planarization\cite{Guan2026}. A large-area graphene sheet grown by chemical vapor deposition (CVD) was transferred from a copper foil onto the substrate and subsequently patterned into contact electrodes.

The graphene contacts serve two distinct functions. First, because the MoSe$_2$ monolayer is electrically grounded through one graphene contact (labeled as the source in Figure~\ref{fig1}), applying a voltage to the back gate generates a vertical electric field that electrostatically dopes the monolayer, analogous to the operation of a conventional field-effect transistor. Second, the graphene contacts are used to apply an in-plane electric field across the monolayer during transport measurements. This configuration enables controlled trion drift while simultaneously allowing electrical characterization of the MoSe$_2$ channel, including its carrier mobility.

To further investigate electrically driven trion transport under a simplified device geometry, we also fabricated Device II, which consists of an intrinsically doped monolayer MoSe$_2$ device without a back gate; its detailed structure is given in the Supplementary Information. In this device, the MoSe$_2$-hBN heterostructure was transferred onto pre-patterned electrodes, and the in-plane electric field was generated directly by the electrodes and coupled to the monolayer MoSe$_2$ through the hBN spacer layer, with almost no current flow. Compared with Device I, Device II eliminates the electrostatic back-gating structure while preserving the capability of applying lateral electric fields to the monolayer.

\section{Trion dynamics measured by optical emission displacement}

For Device I, we excited with a $532~\mathrm{nm}$ green laser and aligned the channel lengthwise along the slit of the spectrometer using a Dove prism. Therefore, we could resolve the steady-state displacement of excitonic complexes by the displacement of their energy-resolved peaks on the spectrometer camera, as shown in Figure~\ref{fig1}(c). The mechanical stability of the setup was characterized independently, confirming that the vibrations introduced an uncertainty in the absolute spatial position of no more than $0.3~\mathrm{\mu m}$. 

\begin{figure}[!htb]
\begin{center}
\includegraphics[width=\linewidth]{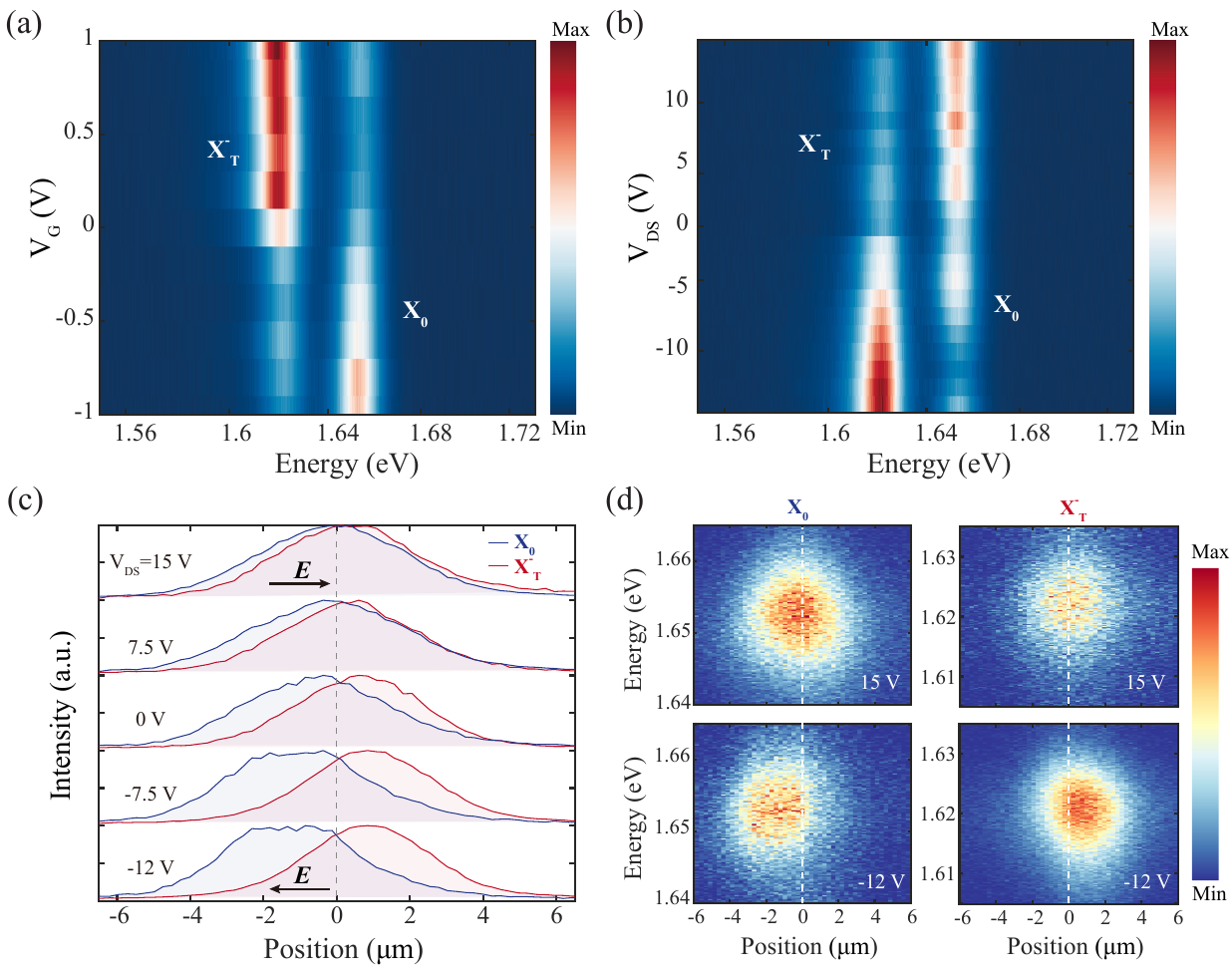}
\end{center}
\caption{Gating-dependent PL emission and trion motion in Device I. (a) PL intensity of the exciton and negative trion as a function of back-gate voltage, for $V_\text{DS} = -12~\mathrm{V}$. As the back gate is increased, the sample becomes increasingly $n$-doped, leading to enhanced trion emission and suppressed exciton emission. (b) PL intensity of the exciton and negative trion as a function of drain--source voltage at a fixed back gate voltage of $V_\text{G} = -0.8~\mathrm{V}$. Although the back-gate voltage is held constant, applying a more negative bias to the source contact also increases electron doping, resulting in a stronger negative-trion emission. (c) Spatial profiles of the exciton and trion emission at different $V_\text{DS}$ under $V_\text{G}=-0.8~\mathrm{V}$. (d) Corresponding CCD images showing the spatially resolved trion and exciton emission under applied drain–source voltages of $V_\text{DS}=15~\mathrm{V}$ and $-12~\mathrm{V}$. The white dashed lines mark the initial excitation position, aligned with the pump laser spot.}
\label{fig2}
\end{figure}
 
Figure~\ref{fig2}(a) and (b) show the trion and exciton photoluminescence (PL) spectral profiles for different back-gate and drain--source voltages, respectively, for Device I. In the gate-dependent PL measurements, the photoluminescence emission peak positions of both the exciton and trion remain nearly unchanged as the carrier density is varied. This weak density dependence is consistent with the picture of a three-body trion bound state rather than with a many-body polaron description\cite{huang2023quantum}. Only the negative trion is observed because monolayer MoSe$_2$ is intrinsically $n$-doped, consistent with previous reports\cite{park2023unveiling}. As expected, increasing the back gate raises the electron density by shifting the Fermi level, as shown in Figure~\ref{fig2}(a). In addition, although the drain-source voltage is primaily applied to generate the in-plane electric field that drives trion transport, it also modifies the carrier density, as evidenced by the evolution of the PL spectra in Figure~\ref{fig2}(b).

Figures~\ref{fig2}(c) and (d) present spectrally resolved real-space PL images of the exciton and trion emission under different drain--source biases. The emission centers of trion and exciton are spatially separated along the direction of the applied electric field. Since sample strain affects both species equally, the relative displacement between the exciton and trion emission can be attributed exclusively to the in-plane electric field.

As expected for negatively charged quasiparticles, the trion emission shifts opposite to the direction of the applied in-plane electric field. Even at zero applied drain--source bias, a small displacement between the exciton and trion emission centers is observed, which we attribute to residual charge redistribution within the device. More remarkably, the exciton emission simultaneously shifts in the opposite direction to the trions, giving rise to a counterpropagating exciton--trion motion shown in Figure~\ref{fig2}(c). Control measurements rule out imaging-system drift as the origin of this effect, confirming that the exciton displacement is intrinsic. Previous spatially resolved studies have reported electric-field-induced trion drift and funneling \cite{Cheng2021,lee2023electric}. In contrast, our measurements reveal a particularly pronounced spatial response that is directly resolved as an increasing separation between the simultaneously imaged exciton and trion PL centers shown in Figure~\ref{fig2}(d). Under the largest applied source--drain bias, the relative separation approaches approximately 1$~\mu\mathrm{m}$, more than three times larger than the experimental positional uncertainty ($\sim0.3~\mu\mathrm{m}$), making the counterpropagating motion clearly visible in the real-space PL images.

\begin{figure}[!t]
\begin{center}
\includegraphics[width=\linewidth]{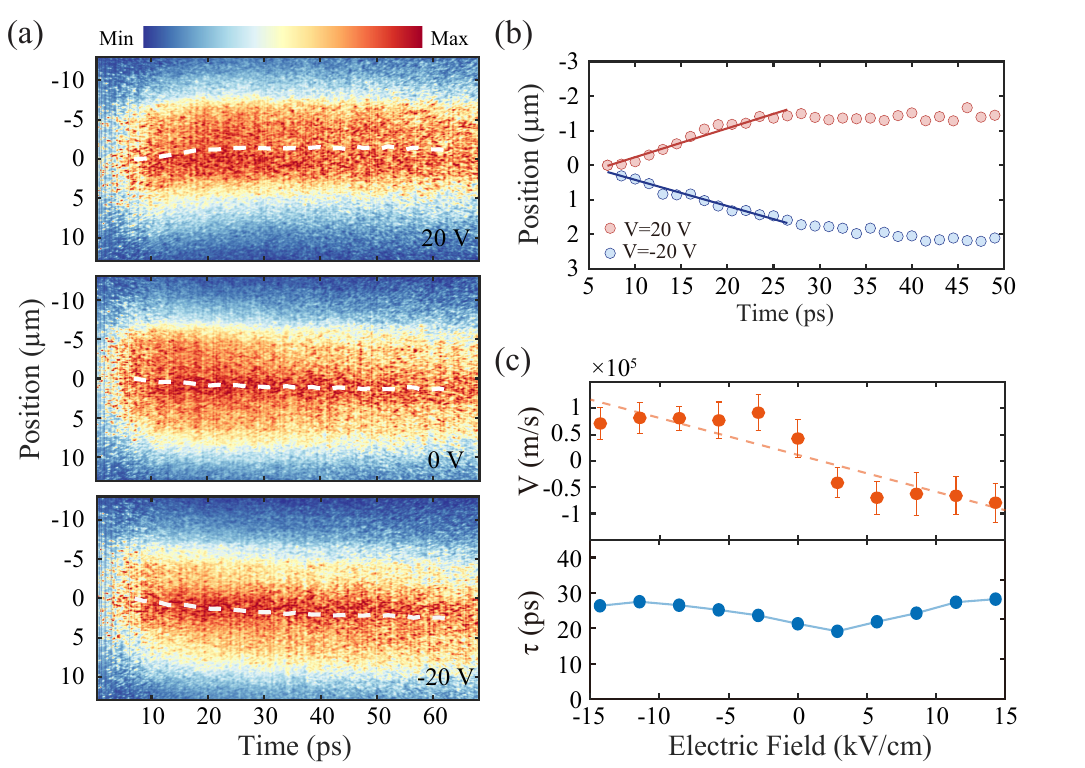}
\end{center}
\caption{Electric-field-dependent real-space transport dynamics of trion emission from Device II. (a) Time-resolved spatial PL maps measured under representative applied voltages of $+20~\mathrm{V}$, $0~\mathrm{V}$, and $-20~\mathrm{V}$. At each delay time, the spatial PL intensity profile is independently normalized to emphasize the temporal evolution of the emission center. The white dashed lines indicate the extracted emission-center positions as a function of delay time, revealing a clear voltage-dependent spatial drift of the trion population. (b) Time-dependent spatial positions extracted under applied voltages of $+20~\mathrm{V}$ and $-20~\mathrm{V}$. The solid lines represent linear fits to the experimental data, from which the trion drift velocities are obtained. (c) Drift velocity (upper) and extracted PL lifetime (lower) as functions of the applied electric field. The orange symbols denote the drift velocities extracted from the linear fits in panel (b). The dashed line shows the corresponding linear fit, yielding an estimated mobility of approximately $700~\mathrm{cm^2/V}$-s for this sample. The blue symbols represent the fitted trion lifetimes, which show only weak voltage dependence over the applied bias range.}
\label{fig3}
\end{figure}

The same effects were also confirmed by our measurements on Device II. For Device II, time-resolved streak-camera measurements were performed using a $532~\mathrm{nm}$ femtosecond pulsed laser with a repetition rate of $80~\mathrm{MHz}$ under relatively large-spot excitation, with a spot diameter of approximately $10~\mathrm{\mu m}$ on the sample. A tunable long-pass filter was used to spectrally isolate the trion emission. As shown in Figure~\ref{fig3}(a), the time-resolved real-space PL maps reveal a clear bias-dependent spatial drift of the trion emission. Within the trion lifetime window, the peak emission locations of the negative trions shift continuously under applied biases of $+20~\mathrm{V}$ and $-20~\mathrm{V}$, respectively, in directions opposite to the corresponding electric fields. The drift direction reverses when the electric-field polarity is reversed. A movie showing the time-resolved drift changing continuously as the voltage is varied is available in the Supplemental Information.

\begin{figure*}[t]
\begin{center}
\includegraphics[width=\textwidth]{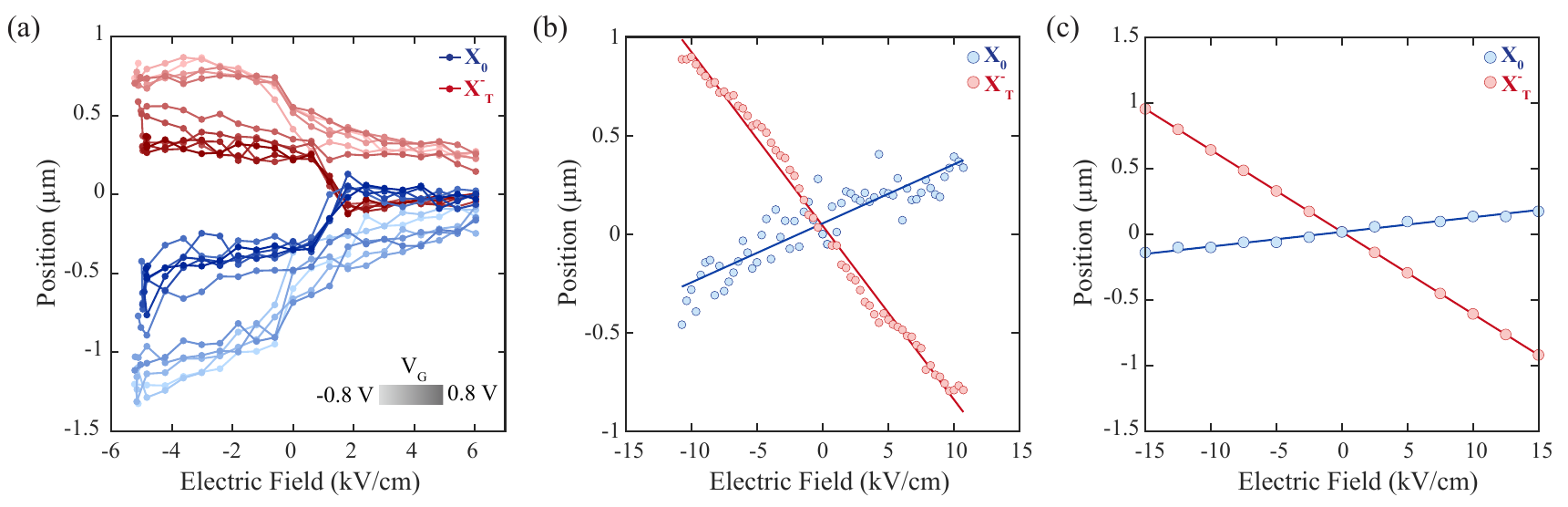}
\end{center}
\caption{Measurement of trion drift and extraction of trion mobility under varying in-plane electric fields. (a) Measured displacements of negative trions (red to blue) and neutral excitons (green to orange) under different back-gate voltages ranging from ($-0.8~\mathrm{V}$  to $+0.8~\mathrm{V}$) and varying in-plane electric fields in Device I. (b) Trion displacement as a function of the in-plane electric field for Device II. A linear fit yields an extracted trion mobility of ($350~\mathrm{cm^2/V}$-s). (c) The resuls of the numerical drift–diffusion simulation of exciton and trion transport presented in the Supplemental Information, which reproduces the experimentally observed linear dependence of the displacement on the applied electric field for both quasiparticle species.}
\label{fig4}
\end{figure*}

To quantify the trion motion, the spatial PL profile at each delay time was fitted with a Lorentzian function to extract the emission center position. The resulting position–time trajectories are plotted in Figure~\ref{fig3}(b). We then fit the early-time data linearly, as indicated by the solid lines, yielding a drift velocity on the order of $10^5~\mathrm{m/s}$ at the highest electric fields. Notably, the spatial drift gradually saturates for delay times longer than approximately $25~\mathrm{ps}$, corresponding to the trion lifetime. Beyond this timescale, a steady state of drift and recombination is reached, and no further measurable drift is observed. The extracted drift velocity and trion lifetime as functions of the externally applied electric field are summarized in Figure~\ref{fig3}(c). The reversal in the sign of the drift velocity directly confirms that the observed motion is governed by the applied electric field. The finite trion drift velocity observed at zero applied voltage, which is slightly beyond the uncertainty of the measurements, may indicate the presence of a residual built-in electric field\cite{zheng2019} that creates a zero-bias drift.

Figure~\ref{fig4}(a) summarizes the electric field dependence of the trions and excitons displacements in Device I, while Figure~\ref{fig4}(b) presents the corresponding measurements for Device II. As seen in Figure~\ref{fig4}(a), the displacement is noticeably asymmetric with respect to the back-gate voltage. We attribute this asymmetry to the higher background electron density at positive gate voltages, which enhances carrier scattering and consequently reduces the mobilities of both trions and excitons. Consistent with the real-space image shown in Figures~\ref{fig2}(c) and (d), and Figures~\ref{fig4}(a) and (b) demonstrate that the electric field drives not only the negatively charged trions but also a simultaneous displacement of the neutral excitons in the opposite direction. This counterpropagating exciton motion suggests a back-action of the drifting trions on the neutral exciton population, as discussed below.

\section{Discussion}

Our measurements directly resolve the electric-field-driven transport of the trions and the accompanying response of neutral excitons in both space and time. Under an applied in-plane electric field, the negatively charged trions experience a net force and acquire a steady-state drift velocity determined by scattering with free carriers and the excitons. 

These measurements also enable an estimate of the trion mobility. Previous attempts at characterizing the mobility of charged excitonic complexes have been limited by either insufficient spatial resolution or the short trion lifetime. Here, we employ two independent ways to estimate the mobility. First, using the time-resolved measurements on Device II, we can directly measure the drift velocity and equate this to the ohmic formula $v_d = \mu E$, where $\mu$ is the mobility. Using the early-time linear fits shown in Figure~\ref{fig3}(b), summarized in \ref{fig3}(c), we obtain a mobility of approximately $700~\mathrm{cm^2/V}$-s. 

The second method combines the measured trion displacement as a function of electric field, shown in Figures~\ref{fig4}(a) and (b), along with the measured lifetime $\tau$ of the trions. This can then be equated to the average drift distance $l = v_d \tau$. A summary of the mobility found this way for Device I is given in the 
{Supplemental Information}. 
At negative gate voltage the mobility is comparable to that of Device II found by direct measurement of the drift velocity, and it drops at positive voltage, consistent with enhanced scattering arising from the higher density of free electrons under stronger $n$-doping.

A simple conceptual model accounts for the counterflow of the trions and excitons. The trions, of course, feel a direct force from the external electric field due to their net negative charge. This creates a gradient of the trion density in the area occupied by the excitons. It is well known that excitons and trions have a repulsive, short-range interaction, akin to the exciton-exciton interaction \cite{Schindler2008,Daniel2021}. Therefore the excitons feel a potential energy proportional to the trion density. Because the excitons feel a gradient of this potential energy due to the gradient of trion density, they feel an effective force in the opposite direction of the trion motion. A modification of the drift-diffusion equation for the two populations is presented in the Supplemental Information. The results of this simulation are shown in Figure~\ref{fig4}(c), which shows that this basic effect can give qualitative agreement with the experimental results.

\section{Outlook and Conclusion}

We have established a direct method for observing trion motion in response to an electric field, enabling a quantitative estimate of the trion mobility. The relatively high mobility we find, {on the order of a few hundred $~\mathrm{cm^2/V}$-s}, may allow practical optoelectronic applications, such as charged-exciton transistors and valleytronic devices \cite{Das2020,Zhang2022}.

Careful measurements of the absolute position show that the excitons are pushed backwards under the conditions when the trions move in response to the electric field. This is surprising at first, because the excitons are charge neutral, but it is well known that trions and excitons repel each other \cite{Rana2020,Katsch2022}, and therefore the gradient of trion density felt by the excitons will push them oppositely to the trions. A simple model, given in the Supplementary Information, accounts for the counterpropagating exciton--trion flow.

This method is general, and therefore is applicable not only to trions, but also to the recently observed quaternions in bilayer structures, which have two free carriers bound to an exciton \cite{sun2021,wan2026}. Observation of quaternion motion under an electric field would give direct confirmation that they are charged bound states. The mobility should be greater than that of the trions, because they have twice the charge, but only approximately 4/3 larger mass. As discussed in earlier work \cite{wan2026}, doubly charged excitons (quaternions) are charged bosons, and therefore are predicted to become a superconductor at low temperature. 

TMDC structures host a rich landscape of excitonic complexes with different charge configurations, together with tunable populations of free carriers. Optical tracking of their motion under applied electric fields provides a unique window into their transport dynamics and many-body interactions, enabling deeper insight into the fundamental physics of excitonic quasiparticles in low-dimensional semiconductors. Future integration with optical microcavities may further extend these studies to the polaritonic regime, opening new possibilities for investigating the transport and many-body physics of hybrid light–matter quasiparticles \cite{Emmanuele2020,Schneider2018}.



\section{Acknowledgments}
This research was supported by the U.S. Army Research Office Grant No. W911NF-24-1-0237. YJ. G and Z.S. were supported by the Innovation Program for Quantum Science and Technology (2023ZD0300300) and Grand NeoBay Development Fund (202509-KW-1.1.1-001). K.W. and T.T. acknowledge support from the CREST (JPMJCR24A5), JST and World Premier International Research Center Initiative (WPI), MEXT, Japan.

\paragraph{Authors contributions:}
Z.S. and D.S. conceived the project and formulated the main experimental concept. D.V., Y.G., Q.W., B.B., A.R., B.V., and J.Y. fabricated the gated TMDC devices. K.W. and T.T. supplied high-quality hBN crystals. J.H. supplied high-quality TMDC crystals. D.V., Y.G., and J.B. designed and carried out the PL measurements. D.V., Y.G., N.Y., Z.S., and D.S. analyzed the data. O.P. performed the numerical simulations in the Supplementary material. All authors discussed the results and contributed to the preparation of the manuscript.

\paragraph{Competing interests:} The authors declare that they have no competing interests. 
\paragraph{Data and materials availability:} All data needed to evaluate the conclusions in the paper are presented in the paper and/or the Supplementary Materials.

\putbib[reference]

\end{bibunit}

\clearpage
\appendix

\begin{bibunit}[apsrev4-2]
\setcounter{table}{0}
\renewcommand{\thetable}{S\arabic{table}}
\setcounter{figure}{0}
\renewcommand{\thefigure}{S\arabic{figure}}
\setcounter{equation}{0}
\renewcommand{\theequation}{S\arabic{equation}}
\setcounter{section}{0}
\renewcommand{\thesection}{S\arabic{section}}
\title{Supplementary Information: Counterdirectional Exciton and Trion Motion in Applied Electric Field}

\maketitle
\section{Additional Data for Device I}

\begin{figure}[H]
	\centering
	\includegraphics[width=0.5\textwidth]{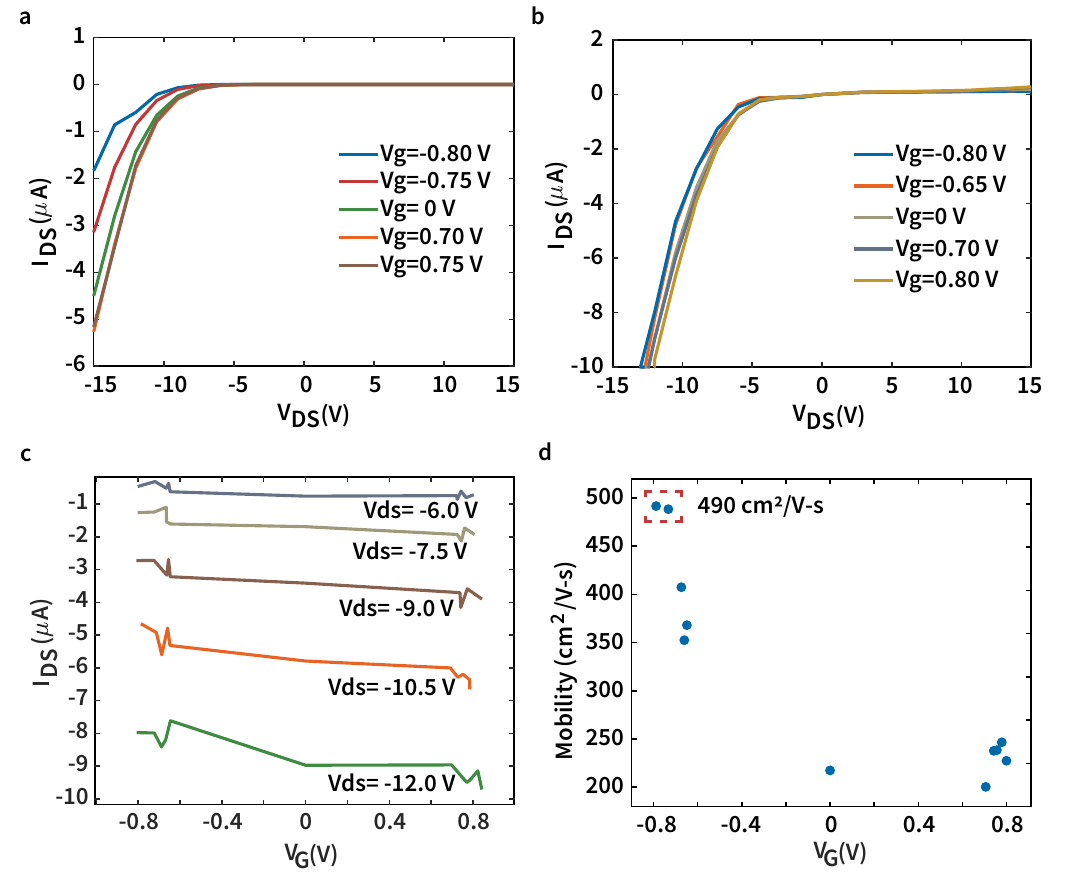}
	\caption{(a) Output ($I_\text{DS}$–$V_\text{DS}$) characteristics measured in the dark, showing linear current–voltage behavior indicative of ohmic transport. (b) Output characteristics under optical illumination, exhibiting a photocurrent approximately twice the dark current at comparable source–drain biases while remaining in the linear transport regime. (c) Transfer characteristics under illumination, showing the dependence of the drain current on gate voltage for source–drain voltages ranging from $-6$ to $-12$ V. (d) Extracted trion mobility as a function of carrier density, with the highest mobility of approximately $490~\mathrm{cm^2/V}$-s obtained at the lowest background electron density.}
	\label{fig.S1}
\end{figure}

Fig.~\ref{fig.S1}(a) and ~\ref{fig.S1}(b) show the $I_\text{DS}$–$V_\text{DS}$ characteristics measured under dark and illuminated conditions, respectively. Under optical excitation, the photocurrent is approximately twice the dark current at comparable source–drain biases, consistent with photocarrier generation in the MoSe$_2$ channel. Despite the relatively high contact resistance, the low-bias $I$–$V$ characteristics remain linear, confirming that the device operates in the ohmic transport regime. The transfer characteristics shown in Fig. ~\ref{fig.S1}(c) demonstrate that the device remains conductive over the entire gate-voltage range investigated ($V_G=-0.8$ to $+0.8$ V) for source--drain voltages between $-6$ and $-12$ V. The measured current remains negative throughout this range, consistent with the intrinsic $n$-type doping of the MoSe$_2$ monolayer. The relatively large contact resistance results in a substantial voltage drop at the graphene/MoSe$_2$ interfaces, thereby limiting the current density in the channel while still allowing reliable electrostatic tuning of the carrier density and establishing the in-plane electric fields required to resolve trion drift. Fig.~\ref{fig.S1}(d) summarizes the extracted trion mobility as a function of carrier density. The highest mobility, approximately $490~\mathrm{cm^2/V}$-s, is obtained at the lowest background doping, after which the mobility decreases with increasing carrier concentration.

\section{Additional Details for Device II}
\begin{figure}[H]
	\centering
	\includegraphics[width=0.5\textwidth]{
    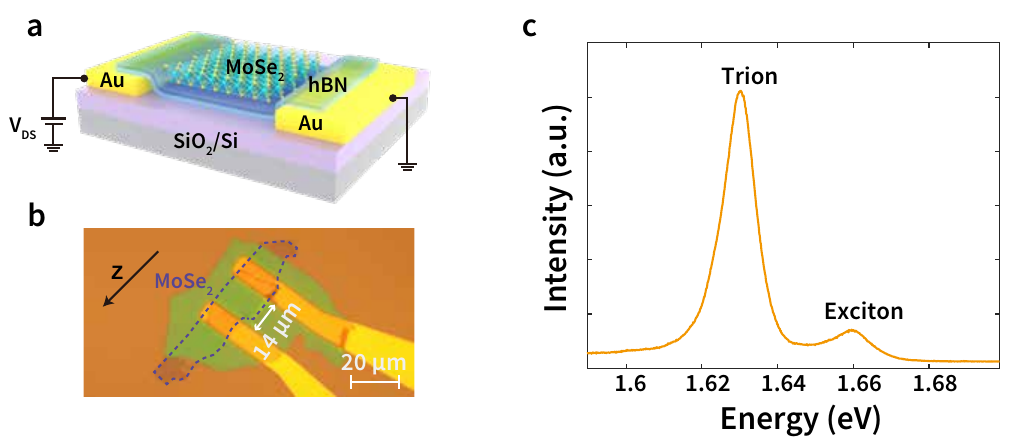}
	\caption{(a) Schematic illustration of the MoSe$_2$-based transistor structure (Device II). (b) Optical micrograph of Device II. The separation between the two gold electrodes is 14 $\mu$m. When a positive bias is applied, the electric field is directed along the Z-axis, as indicated by the black arrow. The dashed region outlines the monolayer MoSe$_2$, which is encapsulated by hBN. (c) PL spectrum of Device II measured at 10 K, showing well-resolved exciton and trion resonances.}
	\label{fig.S2}
\end{figure}

The heterostructure of Device II was fabricated by a dry-transfer method onto pre-patterned electrodes, as shown in Fig.~\ref{fig.S2}(a) and ~\ref{fig.S2}(b). Monolayer MoSe$_2$ and hBN were first mechanically exfoliated from the bulk crystal and then deterministically transferred onto the target substrate. In the assembled device, an hBN flake was used as the dielectric layer. From the optical microscope image, the hBN layer exhibited good uniformity and a clean surface morphology, which is beneficial for applying a spatially homogeneous electric field across the active region and for reducing disorder induced by the surrounding environment.
    
The spacing between the two electrodes was approximately 14 $\mu$m. Also, the V$_{DS}$ is applied between the two electrodes, which is used to apply an electrical field in 2D meterials. As indicated in the figure, the electric field is applied along the $Z$-direction of the device geometry when a positive voltage is applied. This electrode configuration provided a well-defined platform for investigating the optical response of the MoSe$_2$ monolayer under external electrical control.
    
To further evaluate the optical quality of the luminescent medium, photoluminescence (PL) spectroscopy was performed at low temperature. As shown by the PL spectrum measured at 10 K, the exciton and trion resonance peaks can be clearly resolved and are well separated from each other. In addition, no obvious defect-related emission is observed within the measured spectral range. These results indicate that the MoSe$_2$ monolayer possessed good optical quality and a relatively clean emission environment, providing a reliable basis for the following studies of excitonic and trionic optical properties.
	
    All optical measurements for Device II were carried out in a cryogenic system at 10 K. Steady-state PL spectra were measured using a continuous-wave (CW) 532 nm laser as the excitation source. Time-resolved PL measurements were performed using an 80 MHz femtosecond pulsed laser (Coherent Chameleon Ultra II) at 532 nm, and the emission was detected by a streak camera system (Hamamatsu C10910-05). To distinguish different spectral components, a tunable long-pass filter is used to selectively collect the trion emission, whereas a band-pass filter was employed to isolate the exciton emission. Unless otherwise noted, the excitation and collection were performed through the same objective, and the collected signal was guided into the spectrometer for spectral selection before detection.
	
\section{Additional Data for Device II}
    \noindent\textbf{Part 1: Real-space drift distance of excitons and trions as a function of the applied}

\begin{figure}[H]
	\centering
	\includegraphics[width=0.5\textwidth]{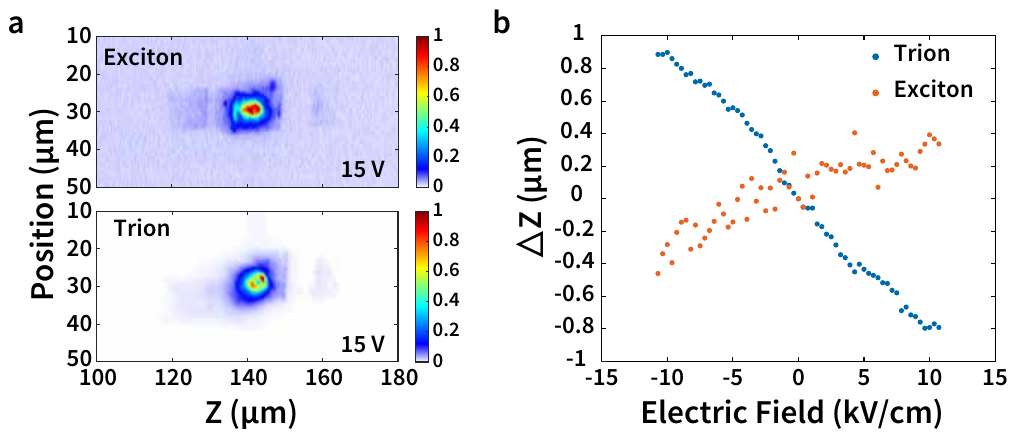}
	\caption{(a) Real-space PL distributions of the exciton (top) and trion (bottom) under an applied voltage of 15 V, for Device II. (b) Voltage-dependent real-space displacement, $\Delta Z$, of the exciton and trion emission spots, for this device. With increasing applied voltage, the exciton and trion shift in opposite directions, indicating opposite spatial responses to the electric field.}
	\label{fig.S3}
\end{figure}

    Fig.~\ref{fig.S3}(a) shows the real-space PL distribution measured under CW excitation at 532 nm. Since the emission profile deviates significantly from a Lorentzian lineshape, conventional fitting methods cannot be reliably applied to determine the emission peak position. Therefore, the peak position is extracted using a center-of-mass method, defined as

\begin{equation}
    Z_{\mathrm{c}}=\frac{\int Z\,I(Z)\,dZ}{\int I(Z)\,dZ}.
\end{equation}

    In the analysis, the PL distribution is first integrated along the vertical direction to obtain a one-dimensional intensity profile. The emission position is then determined using the equation above. The spatial drift distance is defined as $\Delta Z = Z_{\mathrm{c}}(V) - Z_{\mathrm{c}}(0)$, where $Z_{\mathrm{c}}(V)$ is the center-of-mass position at the applied voltage $V$, and $Z_{\mathrm{c}}(0)$ denotes the corresponding position at zero bias.

\begin{figure*}[t]
    \includegraphics[width=\textwidth]{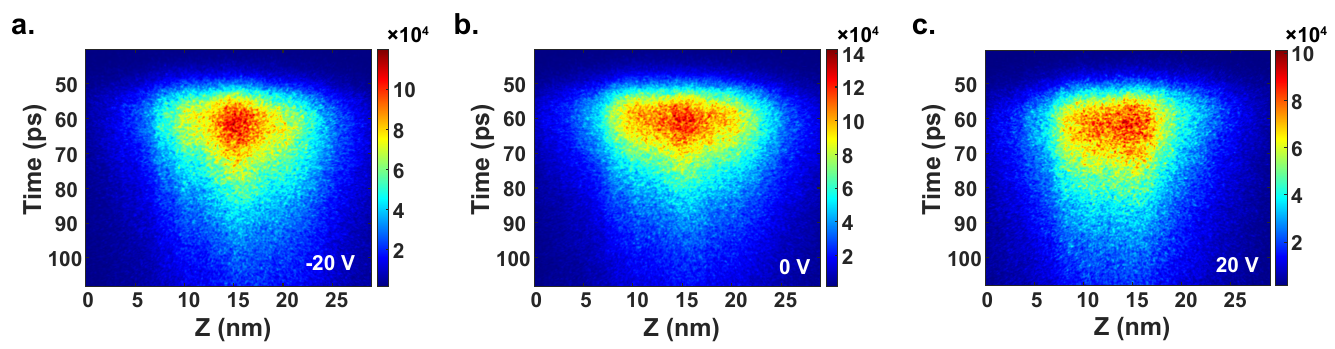}
    \caption{Time-resolved real-space trion PL maps for Device II collected under applied voltages of (a) $-20\,\text{V}$, (b) $0\,\text{V}$, and (c) $20\,\text{V}$. The trion emission exhibits a voltage-dependent spatial drift, with the drift direction opposite to the applied electric field, indicating that the trion is negatively charged. The spatial distribution of the trion emission also evolves over time.}
    \label{fig.S4}
\end{figure*}

\noindent\textbf{Part 2: Time-resolved trion spectra}

    Fig. S4 presents additional time-resolved images of trion transport in Device II under excitation by a laser spot with a diameter of approximately 10 $\mu$m. A movie showing the continuous evolution of the trion drift as the applied electric field is varied is provided as Supplementary Movie 1.

\section{Drift-Diffusion Model Details}
	
    As discussed in the main text, the trions and excitons move in opposite directions under an electric field, as shown, e.g., in Fig. 4(a) and ~\ref{fig.S3}(b). As seen in Fig.~\ref{fig.S3}(b), the motion appears to be linear. In this section, we explain this linear relationship using a drift–diffusion model.

    As discussed in Ref.~\cite{Snoke_2020}, Section 5.8, particles with strong scattering are described by the drift-diffusion equation,
\begin{equation}
    \frac{\partial n}{\partial t} = D\nabla^2 n - \frac{\tau}{m}\vec{\nabla}\cdot(\vec{F}n) + G(\vec{x})-\frac{n}{\tau_L}.
\label{drift-diff}
\end{equation}
    where the scattering time $\tau$ is relation to the mobility $\mu$ and particle mass $m$ by $\mu = e\tau/m$, and the diffusion constant $D$ is related to the mobility by the Einstein relation,
\begin{equation}
    D = \frac{\mu k_BT}{e}.
\label{einsteinD}
\end{equation}
    In Equation (\ref{drift-diff}) we have also added terms for local generation,
    $G(\vec{x})$, and recombination with lifetime $\tau_L$.

    When there are two species, we must write two equations, with repulsions between the species. In principle, we must also allow for interconversion between them, but for our present simple model, we assume the populations are fixed. The repulsive force can be found by writing the interaction potential
\begin{equation}
    U = g(n_X+n_T),
\end{equation}
    where $g$ is the (repulsive) interaction constant between the two species. The force on excitons is then 
\begin{equation}
    \vec{F} = - \vec\nabla U = - g \left(\vec\nabla n_X +  \vec\nabla n_T\right).
\end{equation}
    We then write, for the excitons, 
\begin{eqnarray}
    \frac{\tau}{m}\vec\nabla \cdot (\vec{F}n_X) &=& 
    \frac{\tau}{m}\left[\vec F\cdot\vec\nabla n_X + n_X(\vec\nabla\cdot \vec F)\right] 
\end{eqnarray}
    where $(\tau/m)\vec F$ is the local repulsion velocity, equal to 
\begin{equation}
    \vec{v}_X = -g\frac{\tau}{m}\vec\nabla(n_X+n_T).
\end{equation}
    A similar equation is written for the trions. 
    
    To simulate how excitons and trions move and separate, we considered motion in only one dimension, and mapped the continuous governing equations onto a uniform grid spanning a total length of 40.0 µm with 2048 points. All spatial derivatives were calculated using standard second-order central finite differences. 

\begin{figure}[!t]
	\includegraphics[width=0.5\textwidth]{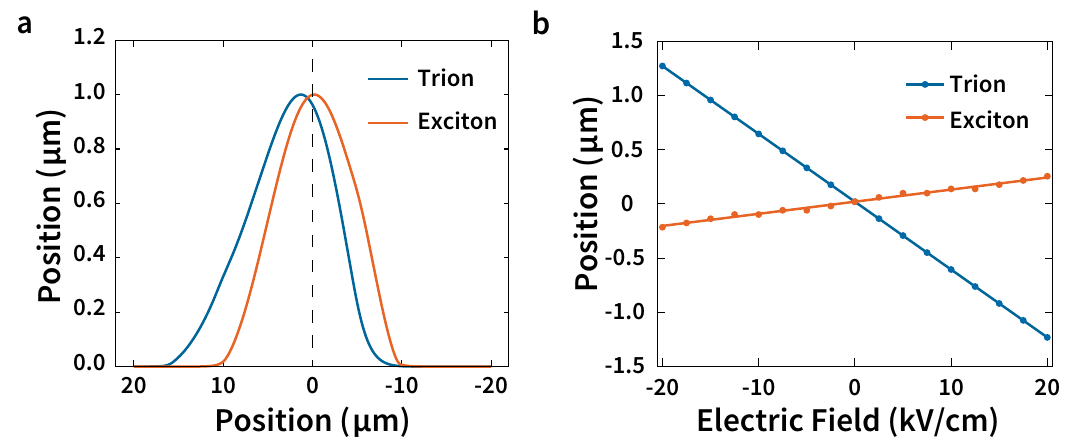}
	\caption{(a) Typical results for the exciton and trion normalized distributions in steady state, for $E = 10$~kV/cm. (b) The exciton and trion peak positions as functions of electric field. The parameters of the model for both figures were generation spot size 3.0~$\mu$m with equal generation rate for both species, exciton and trion lifetimes of 20~ps, trion mobility $\mu = 750$~cm$^2$/V-s, exciton and trion $D = 1.3$ cm$^2$/s (i.e., $k_BT \sim 20$~K), and $gn_0\tau/m = 500$ cm$^2$/s. }
	\label{fig5}
\end{figure}

    Discretizing the space this way turns the equations into a system of 2048 coupled ordinary differential equations (ODEs). Because of the tight grid spacing and the strong nonlinear repulsion terms, this system becomes highly stiff, meaning standard explicit solvers (such as Runge-Kutta) will fail or require impossibly small time steps to remain stable. To solve this efficiently, we used an implicit solver backend (lsode) designed specifically for stiff systems. We started the simulation from an empty initial state and let it evolve over 12.0 ns to guarantee the profiles reached a true steady-state. 
    Fig.~\ref{fig5}(a) shows a typical set of profiles for the excitons and trions, and Fig.~\ref{fig5}(b) shows the positions of the maxima of the two distributions as a function of electric field. This figure corresponds to Figure~4(c) of the main text.

\putbib[SI_reference]

\end{bibunit}
	
\end{document}